\documentstyle[twocolumn,aps,prb,epsfig]{revtex}
\tightenlines
\begin{document}
\twocolumn[\hsize\textwidth\columnwidth\hsize\csname
@twocolumnfalse\endcsname 
\title{Self-consistent local-equilibrium model for density profile and
distribution of dissipative currents in 
 a Hall bar under strong magnetic fields}
\author{Kaan G\"uven and Rolf R. Gerhardts}
\address{
Max-Planck-Institut f\"ur Festk\"orperforschung, Heisenbergstrasse 1,
D-70569 Stuttgart, Germany}
\date{\today}
\maketitle
\widetext
\begin{abstract}
\leftskip 54.8pt
\rightskip 54.8pt
Recent spatially resolved measurements of the electrostatic-potential
variation across a Hall bar in strong magnetic fields, which revealed a clear
correlation between current-carrying  strips and incompressible strips
expected near the edges of the Hall bar, cannot be
understood on the basis of existing equilibrium theories. To explain these
experiments,  we generalize
the Thomas-Fermi--Poisson approach for the self-consistent calculation of
electrostatic potential and electron density in {\em total} thermal
equilibrium to a {\em local equilibrium} theory that allows to treat finite
gradients of the electrochemical potential as driving forces of
currents in the presence of dissipation.
A conventional conductivity model with small values of the longitudinal
conductivity for integer values of the (local) Landau-level filling factor
shows that, in apparent agreement with experiment,
the current density is localized near incompressible strips, whose location
and width in turn depend on the applied current.

\end{abstract}
\leftskip 54.8pt
\rightskip 54.8pt
]
\narrowtext

\section{Introduction}

The question, where the current flows in a Hall bar under quantum-Hall-effect
(QHE) conditions, has been investigated by many authors, and many 
controversial answers have been given. Part of this controversy arises from
the fact that, apart from the total electric current density, different
partial current densities can be defined which integrate to the same total
current. \cite{Komiyama96:2067} Even in the thermodynamic equilibrium state
with vanishing total current, quantum calculations yield finite current
densities which are related to the density variation and the energy dispersion
of the two-dimensional electron gas (2DEG) near the sample edges.
Such current distributions have been calculated using different approaches,
from simple Hartree-type approximations \cite{Wexler94:4815} to sophisticated
treatments of exchange and 
correlation effects within a current-density functional theory
\cite{Geller94:11714,Geller95:14137}.  Unfortunately, there is little
experimental information about the current-density distribution in samples
with zero or small total current. Experiments with high currents, of the order
of the critical current for the breakdown of the QHE, have shown that the
current distribution depends strongly on the mobility of the sample
\cite{Balaban95:R5503}, but also on screening effects caused by nearby
metallic gates \cite{Yahel97:537} or mesoscopic inhomogeneities like arrays of
antidots \cite{Nachtwei97:6731,Nachtwei98:9937}.

Our present work is motivated by a recent series of experiments
\cite{Weitz00:247,Ahlswede01:562,Ahlswede02:165}, in which a scanning force
microscope \cite{Weitz00:349} is used to measure the Hall-potential
distribution across a Hall bar under QHE conditions.  The characteristics of
the potential distribution are found to change drastically with the magnetic
field applied perpendicularly to the sample, i.e., with the Landau level
filling factor $\nu$. While for $\nu$ values far away from integers the
potential varies linearly across the sample (``type I'' behavior) and for
integer and slightly lower 
values a non-linear potential drop in a broad region in the middle of the
sample is observed (``type II''), for slightly larger than integer $\nu$ values
the potential is flat in the center region and
drops across two strips that move with increasing  $\nu$ towards the
sample edges (``type III''). Position and width of these strips coincide
 with those of the incompressible strips \cite{Ahlswede01:562} 
 that are expected to form in the sample due to the
non-linear screening properties of the 2DEG in strong magnetic fields
\cite{Chklovskii92:4026,Chklovskii93:12605,Lier94:7757}.

The fact, that the Hall voltage drops across the incompressible strips,
indicates that the current flows preferably along these strips.
\cite{Wexler94:4815} Such a 
conjecture can be found in an early paper by A.~M.~Chang \cite{Chang90:871}.
Self-consistent equilibrium calculations, which imposed a dissipationless Hall
current as a thermodynamic boundary condition on the 2DEG in a Hall bar,
\cite{Pfannkuche92:7032,Oh97:13519}  could however not confirm this
conjecture. On the contrary, the current-density profile was found to extend
over the whole sample width and to follow closely the electron density
profile.
On the other hand, there is a simple classical argument supporting Chang's
conjecture. If we describe the magnetotransport in the inhomogeneous 
 Hall bar by a local Ohm's law with a position-dependent resistivity
tensor $\hat{\rho}$, we find that the current density is largest where the
longitudinal resistivity is smallest, i.e., along the incompressible strips, if
we assume that $\hat{\rho}(\nu)$ depends on the local filling factor in the
same manner as for a homogeneous sample.
Such a classical local magnetotransport model has previously been used to
explain, for inhomogeneous samples, a finite width of quantum Hall plateaus
without the assumption of localized states, and to simulate the
magnetic-field dependence of Hall-type voltages in Hall bars with internal
contacts.\cite{Woltjer86:149,Woltjer87:104}
More recently such a local model has  also proven useful for
the understanding of the current and electric field distribution in antidot
systems close to the breakdown of the
QHE.\cite{Gerhardts99:2561,Nachtwei98:9937} 

Since the dissipative non-equilibrium current will lead to a
position-dependent electrochemical potential, which changes the
equilibrium electron-density and electrostatic-potential distribution, we have
to generalize the Thomas-Fermi-Poisson scheme for the self-consistent
calculation of the latter to include the current-induced changes. This will be
done in sect.~\ref{model}. Typical results obtained from this approach are
presented in sect.~\ref{results}. In sect.~\ref{conclusions} we will discuss
the most relevant results in the light of the motivating experiments
\cite{Ahlswede01:562}, and we will comment on apparent limitations of our
model.  
Some preliminary results of this work have been presented
previously.\cite{Guven02:Oxford}

\section{Model} \label{model}
Following previous work \cite{Chklovskii93:12605,Oh97:13519}
we model the Hall bar as a 2DEG in the
plane $z=0$, being restricted to the strip $|x|<d$ and translation invariant
in the $y$ direction. Physically, the confinement potential of the 2DEG is
produced by a homogeneous background charge in the strip, so that the 
density of free charges has the form $\rho(x)\delta(z)$ with
\begin{equation} \label{sp-charge}
\rho(x)=e[n_0 -n_{\rm  el}(x)]\,\theta(d^2-x^2) \,,
\end{equation}
with $n_{\rm  el}(x)$ and  $n_0$ the surface densities of 2DEG and
background, respectively. Electrostatic boundary conditions are fixed by the
assumption of metallic halfplanes of constant potentials $V_L$ and $V_R$ in
$z=0$, $x<-d$, and in $z=0$, $x>d$, respectively, and by dielectric constants 
$\kappa_{>}$ and  $\kappa_{<}$ in the half spaces $z>0$ and $z<0$. With these
assumptions, for $z \neq 0$ 
the electrostatic potential $V(x,z)={\rm Im} F(\zeta)$ satisfies
 the Laplace equation and can be written as the imaginary part of an
analytic function $ F(\zeta)$ of the complex variable $\zeta=x+iz$, which is
determined by its boundary conditions on the real
axis.\cite{Glazman91:32,Chklovskii92:4026,Chklovskii93:12605}

We shall recall briefly the use of this electrostatic model for the
description of 
the thermodynamic equilibrium state \cite{Lier94:7757,Oh97:13519} and then
propose an extension to current-carrying stationary non-equilibrium states.

\subsection{Thermal equilibrium}
To determine $n_{\rm el}(x)$ we need the electrostatic potential
$V(x)=V(x,z=0)$ in $|x|\leq d$. It can be written as the sum of three terms,
\begin{equation} \label{total-pot}
V(x)=V_0(x)+V_{\rm g}(x)+ V_H(x)\,,
\end{equation}
where
\begin{equation} \label{gate-pot}
V_{\rm g}(x) = \frac{V_L+V_R}{2} + \frac{V_R-V_L}{\pi} \, \arcsin
  \left(\frac {x}{d}\right)
\end{equation}
is determined by the potential values on the in-plane gates and accompanied
by compensating induced charges on the gates. The free charges determine
$V(x,z)={\rm Im}F(\zeta)$ with \cite{Oh97:13519} [we write the potential as
potential energy of an electron, i.e., including a factor $-e$]
\begin{equation} \label{complexF}
\frac{dF}{d\zeta}=\frac{i}{\pi w(\zeta)} \int_{-d}^d \! dx \,
\frac{\sqrt{d^2-x^2}}{\zeta -x}\, \frac{2\pi e}{\bar{\kappa}}\,\rho(x) \,,
\end{equation}
 where $ w(\zeta)=\sqrt{d^2-\zeta^2}$ is analytic  except on
 cuts at $z=0$, 
 $|x|\geq d$, and $\bar{\kappa}=(\kappa_{>} +\kappa_{<})/2$. By integration one
 obtains the (bare) confinement potential for $ n_{\rm  el}(x)\equiv 0$ as
\begin{equation} \label{backgr-pot}
V_0(x)= -E_0 \, \sqrt{1-\left(\frac{x}{d}\right)^2} \,, \qquad E_0=\frac{2\pi
  e^2}{\bar{\kappa}} \, n_0 d\,.
\end{equation}
The Hartree potential due to the 2DEG follows as
\begin{equation} \label{Hartree-pot}
 V_H(x)=\frac{E_0}{\pi n_0 } \int_{-1}^1 \! d\xi '\,
  K(\xi,\xi')\, n_{\rm  el}(\xi'd)\,,
\end{equation}
with the kernel \cite{Oh97:13519}
\begin{equation} \label{kernel}
 K(\xi,\eta)= \ln \left| \frac{ \sqrt{
  (1-\xi^2)(1-\eta^2)}+1- \xi \eta}{\xi -\eta} \right| \,.
\end{equation}

The electron density is, in turn, determined by the potential $V(x)$. We
assume that 
$V(x)$ varies slowly on the scale of typical quantum mechanical lengths,
notably the magnetic length $l_m=\sqrt{\hbar/m\omega_c}$ defined by the
cyclotron frequency  $\omega_c=eB/mc$, and calculate the electron density in
the Thomas-Fermi approximation 
\begin{equation} \label{TFA}
n_{\rm  el}(x) = \int dE\, D(E) \, f\big( E+V(x)-\mu^* \big) \,,
\end{equation}
where $f(E)= 1/[1+\exp (E/k_BT)]$ is the Fermi distribution and $D(E)$ the
(single particle) density of states (DOS) of the 2DEG. For the Hall bar in the
absence of a magnetic field, $B=0$,  we take  $D(E)=D_0 \, \theta(E)$, with
$D_0=m/(\pi\hbar^2)$, while for large $B$ we will use a (suitably broadened)
Landau DOS.

Solving Eqs.~(\ref{total-pot}) and (\ref{TFA}) self-consistently for {\em
  constant} electrochemical potential $\mu^*$, we obtain the electron density
  and the electrostatic potential in the thermal equilibrium state.
The value of $\mu^*$ determines the average electron density and vice versa
  (for fixed $T$, $B$, etc.).

\subsection{Local equilibrium with imposed current}
If a stationary net current is imposed on the Hall bar, this leads to
position-dependent  
current densities and electric fields which, in the linear response regime,
are interrelated by Ohm's law. The relevant field driving the net current is
the gradient of the electrochemical potential, ${\bf \nabla}\mu^*$, which
vanishes in thermal equilibrium. In view of Eq.~(\ref{TFA}), we should expect
that a position-dependent $\mu^*$ will lead to a modified
electron density $n_{\rm el}(x)$ and this, according to Poisson's equation, to
a modified electrostatic potential. Our aim is to extend the self-consistent 
Thomas-Fermi-Poisson approximation for the thermal equilibrium state to the 
stationary non-equilibrium situation defined by the imposed current. To do so,
we adopt the widely used assumption of {\em local equilibrium}: thermodynamic
variables are assumed to vary only slowly in space and to satisfy locally the
same relations as they would do in a homogeneous thermodynamic equilibrium
state. Then, for a given position-dependent $\mu^*(x)$, we can use again
Eqs.~(\ref{total-pot}) and (\ref{TFA}) to calculate $V(x)$ and 
$n_{\rm el}(x)$.

We will now formulate a model that allows to
calculate ${\bf \nabla}\mu^*$, and thus the position-dependent electrochemical
potential up to a constant, provided the electron density $n_{\rm el}(x)$ and
the total current
\begin{equation} \label{Itot}
I=\int_{-d}^d dx\, j_y(x,y)
\end{equation}
are given. In the spirit of the Thomas-Fermi and the local equilibrium
approximation, which  assume that electrostatic potentials and thermodynamic
variables vary on a scale much larger than quantum lengths, we also assume
that the current density $ {\bf j}({\bf r})=\big(j_x(x,y),j_y(x,y)\big)$
 and the electric field satisfy the local version of Ohm's law,
\begin{equation} \label{local-Ohm}
 \hat{\rho} ({\bf   r})\, {\bf j}({\bf r})={\bf  E}({\bf
 r})\equiv  {\bf \nabla} \mu^* ({\bf   r}) /e \,,
\end{equation}
where the resistance tensor $ \hat{\rho} ({\bf   r})=\big[ \hat{\sigma}\big(
n_{\rm el} ({\bf   r})\big)\big]^{-1}$ is assumed to depend on position only
via the electron density $n_{\rm el} ({\bf   r})$.

Assuming a stationary situation with translation invariance in $y$ direction
(i.e. $ \hat{\rho}$, $ {\bf j}$, and ${\bf  E}$ are independent of $y$),
we obtain from the Maxwell equations $ {\bf \nabla} \cdot  {\bf j}({\bf r}) =
0$ and ${\bf \nabla} \times  {\bf  E}({\bf   r})= {\bf 0}$ that the 
current density $j_x$ across  and the field $E_y$ along the bar are
independent of $x$,
\begin{equation} \label{jE1}
j_x \equiv 0 \,, \quad E_y(x)\equiv E_y^0 \,.
\end{equation}
With
$\rho_{yy}=\rho_{xx}=\rho_l(x)$, $\quad \rho_{xy}=-\rho_{yx}=\rho_H(x)$ we
further get
\begin{equation} \label{jE2}
 j_y(x)=\frac{1}{\rho_{l}(x)}\,E_y^0\,, \quad
E_x(x)=\frac{\rho_H(x)}{\rho_l(x)}\,E_y^0 \,.
\end{equation}
From these results we obtain
\begin{equation} \label{muxy}
\mu^*(x,y)=\mu^*_0 +e E_y^0 \,\left\{ y+ \int_0^x dx' \,
\frac{\rho_H(x')}{\rho_l(x')} \right\} \,, 
\end{equation}
where $\mu^*_0$ occurs as an undefined constant,
and from Eq.~(\ref{Itot}) we get the normalization
\begin{equation} \label{normalize}
E_y^0=\frac{I}{ \int_{-d}^d dx \,[1/\rho_{l}(x)]}\,.
\end{equation}

The new self-consistency problem for the stationary state is completely
defined, if we choose a model for the dependence of the conductivity tensor on
the electron density. To solve it iteratively, we start with zero current,
$I=0$, and solve the old equilibrium problem. Then we take a fixed value $I
\neq 0$ and  proceed as follows.  

In the next step of the new iteration, we use $n_{\rm el}(x)$ from the
previous step and calculate for the given 
$I$ the electrochemical potential (\ref{muxy}). Then we put this into
Eq.~(\ref{TFA}). To compensate the $y$-linear term, we add an identical term
to the electrostatic potential. This guarantees that the electron density
remains independent of $y$. Then we solve the ``old'' problem,
Eqs.~(\ref{total-pot}) and (\ref{TFA}) with the modified $x$-dependent
electrochemical potential, self-consistently to determine $V(x)$ and  
$n_{\rm el}(x)$, choosing $\mu^*_0$ so that the average electron density
remains the same as without current. Convergence of the ``old'' problem
completes this step of the new iteration. The steps of the new iteration
are repeated until there is practically no further change of  $n_{\rm el}(x)$.

We have performed these self-consistent calculations for two types of
electrostatic boundary conditions. First we started with a symmetric electron
profile and $V_G=V_R-V_L=0$, and assumed that an applied current does not
change the potentials of the in-plane gates. Then, in a strong magnetic field
and under an imposed current, we obtain an asymmetric electron profile that is
shifted towards one of the sample edges. This is an obvious consequence of the
Lorentz force on the drifting electrons. In this situation, the applied
current leads to a change of the induced charges in the in-plane
gates.

Since $\partial V(x,z)/\partial z= {\rm Re}[dF/d\zeta]$, we can use
Eq.~(\ref{complexF}) to calculate the induced charge density in the in-plane
gates, $\rho_{\rm ind}(x)=[\bar{\kappa}/(2\pi e)]{\rm Re} F'(x+i0^+)$.  For
the total induced charge $Q_R=\int_d^{\infty}dx \rho_{\rm ind}(x)$ 
in the right gate  this yields
\begin{equation} \label{QR}
\frac{Q_R}{e n_0 d}=-\frac{2}{\pi} \int_{-1}^1 \! d\xi 
 \left[1-\frac{ n_{\rm  el}(\xi d)}{n_0}\right]
 \arctan \sqrt{\frac{1+\xi}{1-\xi}}\,.
\end{equation}
The corresponding result $Q_L$ for the left gate is obtained from
Eq.~(\ref{QR}) by replacing $\xi$ under the square root by $-\xi$. The sum of
these induced charges compensates the free charges, $Q_R+Q_L=-\int_{-d}^d dx
\rho(x)$, and their difference vanishes if the electron density is symmetric, $
 n_{\rm  el}(-x)= n_{\rm  el}(x)$.

The asymmetric density profile, resulting for the current-carrying stationary
state from the requirement $V_G=0$, leads to $Q_R \neq Q_L$. Since $Q_R+Q_L$ is
kept constant, this means that  charge must flow from one  gate to the other as
the stationary state is established.
As an alternative electrostatic boundary condition, which may be more realistic
in certain situations,
we investigated   the ``floating gate'' condition, requiring that $Q_R$ (and
thus $Q_L$) is kept constant and a finite voltage between the in-plane gates
builds up.

\section{Results} \label{results}
\subsection{Classical regime}
In the classical regime of low magnetic field and high temperature, the
magnetic field should not affect the thermodynamic equilibrium state
\cite{Gossmann98:1680} (Bohr--van-Leeuwen theorem). Therefore, we use the
simple DOS  $D(E)=D_0 \, \theta(E)$, which for zero temperature renders
Eq.~(\ref{Hartree-pot}) into 
\begin{equation} \label{linintgl}
V_H(\xi d)=\frac{1}{\alpha} \int_{\beta_L}^{\beta_R} d\xi' \, K(\xi,\xi') \big[
E_F- V(\xi'd)\big] \,,
\end{equation}
where the dimensionless parameters $\beta_L<\beta_R$ define the edges of the
density profile, $V(\beta_L d)=V(\beta_R d)=E_F=\mu^*(T=0)$, and where
\begin{equation} \label{alpha}
\alpha=\pi n_0/(E_0 D_0)=\pi a_0/d \,,
\end{equation}
with $a_0=\bar{\kappa}\hbar^2/(2m e^2)$ the screening length. [Due to a
misprint, in the denominator in Eq.~(15) of Ref.~\onlinecite{Oh97:13519} the
factor $\pi$ is missing.] Together with Eq.~(\ref{total-pot}),
Eq.~(\ref{linintgl}) represents a linear integral equation for $V(\xi d)$ in
the interval $\beta_L\leq \xi \leq \beta_R$, which can easily be solved
numerically. Solutions of this linear equation are used as  starting points
for all numerical calculations at finite temperature and magnetic field, which
lead to non-linear integral equations that must be solved iteratively.

As has been shown in Ref.~\onlinecite{Oh97:13519}, with decreasing values of
$\alpha$ screening becomes more effective, and a voltage $V_G$ applied across
the bar leads to a shift and deformation of the equilibrium electron density
profile $n_{\rm el}(x)$. At finite temperature, the sharp edges of the
zero-temperature profiles are smeared out.

In the non-equilibrium calculations, we use the Drude model for the
resistivity tensor,
\begin{equation} \label{Druderho}
\rho_l(x)=1/\sigma_0(x), \quad \rho_H(x)=\omega_c\tau\, \rho_l(x),
\end{equation}
with $ \sigma_0(x)=(e^2\tau/m) n_{\rm el}(x)$. 
Then, according to Eq.~(\ref{jE2}), the current density $j_y(x)= \sigma_0(x)
E_y^0$ is proportional to the electron density, and $E_x(x)=\omega_c\tau\,
E_y^0$ is constant. Strictly speaking the last result, which follows from the
fundamental 
linear response equation $j_x(x)=\sigma_{xx}(x)[E_x-\omega_c\tau\,E_y]\equiv
0$, can be justified from our local approach only at positions $x$ where $
n_{\rm el}(x) \neq 0$. 
On the other hand, within our local equilibrium model we expect the
electrochemical potential to be constant in regions where no (dissipative)
current flows and where no electrons are. Therefore we put $E_x(x)=0$ if
$n_{\rm el}(x) =0$. In practical calculations we use
\begin{equation} \label{Ex_model}
E_x(x)=\omega_c\tau\,E_y^0\, \theta\big(n_{\rm el}(x)/n_0 -\epsilon \big) \,
\end{equation}
with $\epsilon=10^{-4}$, which defines effective edges $b_L<b_R$ of the
density profile by $n_{\rm el}(b_L)=n_{\rm el}(b_R)=\epsilon n_0$.

\begin{figure}[h]  \centering
\noindent  
\includegraphics[width=1.05\linewidth]{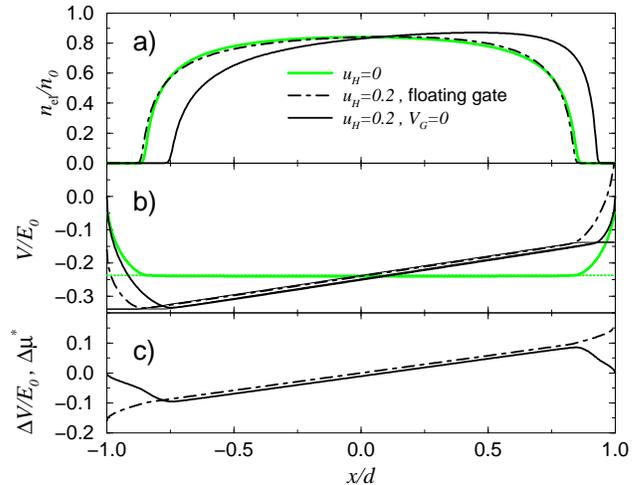}
\caption{\small a) Density profile $n_{\rm el}(x)$ and b) total potential
  $V(x)$ in equilibrium (gray lines) and with current $I\propto u_H$ (black
  lines) for fixed
  gate voltage $V_R=-V_L=0$ (solid) and for floating gates (dash-dotted
  lines), and c) induced Hartree potentials $V(x;I)-V(x,0)$ for both
  cases. The thin lines indicate the electrochemical potentials
  $\mu^*(x)$. $\alpha=0.01$, $\beta_R=-\beta_L=0.848$. 
\label{fig:drustrom}}
\end{figure}

 Typical results are shown in Fig.~\ref{fig:drustrom}, where for the
 current-carrying states the variation of
 the electrochemical potential, $\Delta \mu^*\equiv \mu^*(b_R)-\mu^*(b_L)
=(b_R-b_L) \omega_c\tau E_y^0$, is fixed as $u_H= \Delta \mu^* /E_0=0.2$. 
This yields the current $I=u_H\,[2d/(b_R-b_L)]\,(e^2\bar{n}_{\rm
 el}/m\omega_c)\, E_0/e$, where $ 2d\, \bar{n}_{\rm el}=\int_{-d}^d dx \,n_{\rm
 el}(x)$ is fixed, $\bar{n}_{\rm el}/n_0=0.6278$. 
We observed that, for fixed vanishing gate voltage and
$0<u_H\lesssim 0.2$ the shift (and deformation) of the
 density profile increases roughly linearly with $u_H$.
Under floating gate
 conditions, for $u_H=0.2$ a voltage $V_R-V_L=0.319 E_0$ builds up between the
 in-plane  gates. This is larger than the linear extrapolation
 $u_H\,[2d/(b_R-b_L)] E_0$ 
 due to the singular slopes of the self-consistent potential at the edges of
 the gates (see Fig.~\ref{fig:drustrom}).
Figure~\ref{fig:drustrom} shows that, over the whole range of finite density,
 the  self-consistently calculated total potential $V(x)$ follows very closely
 the  linear position-dependence of the electrochemical potential.  This
 observation holds for both the fixed-gate-voltage and the floating-gate
 boundary conditions. In the following we will consider only the more
 realistic floating-gate  boundary conditions.

\subsection{Quantum regime}
For strong, quantizing magnetic fields we should use a suitable form of a
broadened Landau DOS
\begin{equation} \label{landau-dos}
 D(E)=\frac{g_s}{2\pi l_m^2}\sum_{n=0}^{\infty} \, A_n(E)\,,
\end{equation}
where $ A_n(E)$ is the spectral function of the $n$-th Landau level with
 energy eigenvalue  $\varepsilon_n=\hbar\omega_c (n+1/2)$,  and
$g_s=2$ accounts for spin degeneracy.
The model to be used for the resistivity tensor should show the characteristic
behavior known from the quantum Hall regime, notably (nearly) vanishing
$\rho_l(x)$ at (even) integer local filling factor $\nu(x)=2\pi l_m n_{\rm
  el}(x)$. Moreover, the approximations for the conductivity tensor and the DOS
should satisfy certain consistency relations (consequences of the equation of
continuity).\cite{Gerhardts75:285} However, before we address such
 sophisticated questions we want to present a very simple model.
\subsubsection{Simplified model}
First we follow previous work \cite{Lier94:7757,Oh97:13519} and consider the
bare Landau DOS, taking in Eq.~(\ref{landau-dos}) $
A_n(E)=\delta(\varepsilon_n-E)$.
This leads, in thermal equilibrium, to the appearance of incompressible strips
of finite width at 
integer values of the local filling factor. The temperature dependence of
these strips has been discussed in Ref.~\onlinecite{Lier94:7757} and 
the dependence of position and width of the strips on magnetic field and
applied gate voltage $V_R-V_L$ has been investigated in
Refs.~\onlinecite{Oh97:13519} and \onlinecite{Oh97:108} for the present Hall
bar geometry. 
To investigate the effect of an imposed current, we first use a simplistic
model for the resistance tensor, that has
been used successfully for the 
calculation of the current density in an antidot system in a strong magnetic
field. \cite{Gross98:60,Gerhardts99:2561} For
$\sigma_{yx}(x)=-\sigma_{xy}(x)=\sigma_H(x)$ we take
\begin{equation} \label{gross_H}
\sigma_H(x)=(e^2/h)\, \nu(x) \,,
\end{equation}
which yields the correct values at integer filling factors, but no quantum
Hall plateaus. To simulate the behavior of $\sigma_{xx}=\sigma_{yy}=\sigma_l$
near integer filling $\nu=2$, we approximate  \cite{Gerhardts99:2561} the
longitudinal conductivity as
\begin{equation} \label{gross_l}
\sigma_l(x)=\sigma_H(x)\,\big[\epsilon + \big(2-\nu(x)\big)^2 /4\big] \,,
\end{equation}
\begin{figure}[h]  \centering
\noindent  
\includegraphics[width=1.05\linewidth]{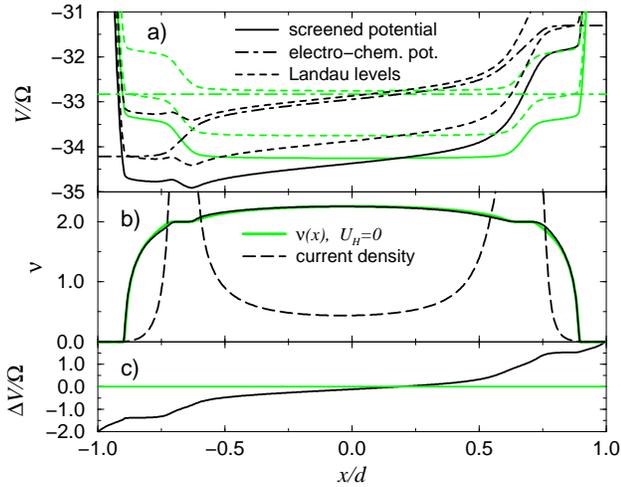}
\caption{\small a) Self-consistent potential $V(x)$  and b) normalized
 density $\nu(x)$ for zero (thick solid gray) and finite (thick solid black
 lines)  current, 
 calculated for model (\protect\ref{gross_l}) with $\epsilon=0.002$. 
 a) also shows the corresponding $\mu^*(x)$ (dash-dotted) and the Landau
 levels $V(x)+\Omega(n+1/2)$ (dashed lines), and b) the current
 density (long-dashed line, arbitrary units). c) shows the
 current-induced change of the self-consistent potential (black). Model
 parameters (see text) are $\alpha=0.02$, $\Omega \equiv \hbar
 \omega_c=E_0/200$, $k_BT/\Omega=0.04$, $U_H \equiv \Delta \mu^*=3 \Omega$.
\label{fig:flgsimpl}}
\end{figure}
\noindent
with a small but finite positive value $\epsilon$ ($\sim 10^{-3}$) to avoid
divergencies.  This describes correctly that $\rho_l(x) \propto \sigma_l(x)$
becomes very small at the incompressible strips with local filling factor
$\nu(x)=2$, although the analytical dependencies for $\nu \neq 2$ are not
correct (see Fig.~\ref{fig:slongvsnu}b below). 

Nevertheless, this simple model is able to reproduce characteristic
features observed in the experiment \cite{Ahlswede01:562} as is shown in
Fig.~\ref{fig:flgsimpl}.  The current flows preferably along the
incompressible strips, 
 where  $\nu(x)=2$ and the longitudinal resistivity is smallest, and there 
 the gradient of $\mu^*(x)$ is largest. Due to  the 
 Thomas-Fermi-Poisson self-consistency requirement, the total potential $V(x)$
 is forced to follow $\mu^*(x)$ closely, so that the current-induced change of
 the electron density profile is small (which keeps the change of
 electrostatic energy small). The induced $\Delta V(x)$ follows closely
 $\mu^*(x)$ and varies mainly in the region of incompressible strips.

\subsubsection{Gaussian level broadening}
As a more  realistic model for the longitudinal conductivity  we use the
Gaussian model \cite{Gerhardts75:285} 
\begin{equation} \label{gauss-sigl}
\sigma_l=\frac{e^2 g_s}{h}\!
\int_{-\infty}^{\infty}\!\!dE\left[\!-\frac{df}{dE}\!\right]\! \sum_{n=0}^{\infty}\!
\Big(n\!+\!\frac{1}{2}\Big) \big[\sqrt{\pi}\,\Gamma A_n(E) \big]^2
\end{equation}
with the spectral function
\begin{equation} \label{gauss-spec}
A_n(E)= \frac{\exp(-[\varepsilon_n-E]^2/\Gamma ^2)}{\sqrt{\pi}\, \Gamma}\,,
\end{equation}
\begin{figure}[h]  \centering
\noindent  
\includegraphics[width=1.05\linewidth]{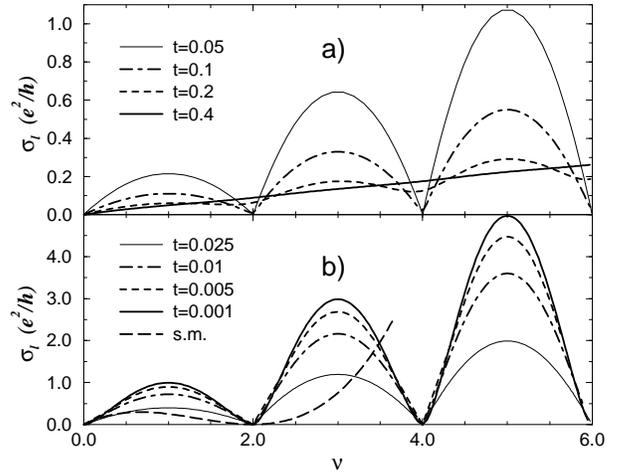}
\caption{\small Longitudinal conductivity $\sigma_l$ versus filling factor
  $\nu$ for the Gaussian model  (\protect\ref{gauss-sigl})
with $\Gamma/\hbar \omega_c=0.035$ for (a) high
  and (b) low values of the reduced temperature $t=k_BT/\hbar \omega_c$. The
  thick solid curves coincide with the limits for high temperature
  [$k_BT>\hbar \omega_c$, (a)] and zero temperature [$k_BT \ll \Gamma$, (b)].
The long-dashed curve in (b) indicates the model
  (\protect\ref{gross_l}). 
\label{fig:slongvsnu}}
\end{figure}
\noindent
which then, for consistency reasons, should also be used in the DOS,
Eq.~(\ref{landau-dos}). An alternative, which leads to qualitatively the same
 results, would be to use the
self-consistent Born approximation \cite{Ando82:437,Woltjer87:104} which would
replace the 
normalized Gaussians by normalized half-ellipses. To avoid divergencies, we
replace $\sigma_l$ of Eq.~(\ref{gauss-sigl}) for
$\nu >1$ by  $\max (\sigma_l,\sigma_H/10^4 )$.

Together with $\nu=2\pi l_m^2 \int \!dE \,f(E-\mu)D(E)$,
Eqs.~(\ref{landau-dos}),  (\ref{gauss-sigl}), and (\ref{gauss-spec}) can be
used to calculate $\sigma_l$ as function of $\nu$. Results are plotted in
Fig.~\ref{fig:slongvsnu} for several temperatures. For high temperatures,
$k_BT \gtrsim 0.3\,\hbar \omega_c$, one gets the modified Drude result,
$\sigma_l=\sigma_H/(\omega_c \tau_{\rm gauss})$ with $\sigma_H=(e^2/h)\,\nu$
and $\hbar/ \tau_{\rm gauss}=\sqrt{\pi/2}\,\Gamma$.

\begin{figure}[h]  \centering
\noindent  
\includegraphics[width=1.0\linewidth]{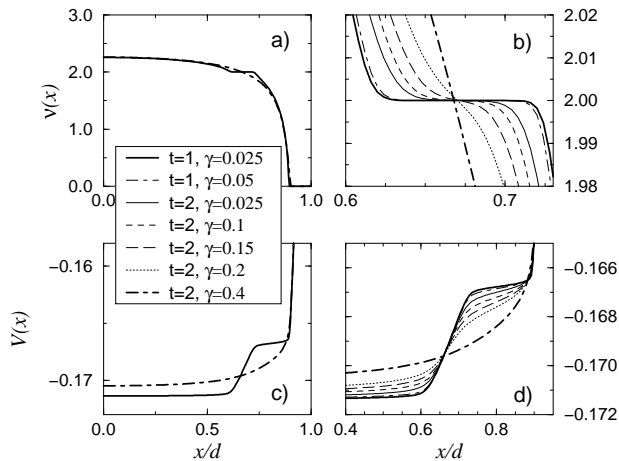}
\caption{\small Density profile [a), b)] and potential [c), d)] calculated
  with a Gaussian DOS; b) and d) show results in the region of the
  incompressible strip  for several
  values of temperature, $k_BT/\hbar \omega_c =t/50$, and level
  broadening, $\gamma=\Gamma / \hbar \omega_c$. The thick curves are plotted
  in a) and c) for one half of the symmetric sample ($\alpha=0.02$, $\hbar
  \omega_c/E_0=0.005$). 
\label{fig:gaussstrips}}
\end{figure}

To proceed, we first investigate the effect of the Landau-level broadening
$\Gamma$ on the existence and width of the incompressible strips.
Results of self-consistent calculations are shown in
Fig.~\ref{fig:gaussstrips}. The width of the incompressible strips shrinks
with increasing temperature and with increasing level broadening. For
$k_BT/\hbar \omega_c \lesssim 0.04$ and $\Gamma/\hbar \omega_c  \lesssim 0.1$
clearly visible incompressible strips exist. Thus, collision broadening of the
Landau DOS does not change the screening properties of the 2DEG qualitatively,
provided the width of the Landau levels remains small enough as compared with
the cyclotron energy.

Next we perform the self-consistent calculation of the charge and current
densities and of the electrostatic and electrochemical potentials for the
Gaussian model. To 
achieve convergence of the nested self-consistency loops for given values of
temperature $T$, cyclotron energy $\Omega\equiv \hbar\omega_c$ and total
current $I \propto U_H \equiv \Delta \mu^{\star}$, we proceed as follows. 
First we define the density profile by solving the linear integral
equation [Eqs.~(\ref{total-pot}) and (\ref{TFA})] for $T=0$, $B=0$ and
$I=0$. Then we raise, still for  $B=0$ and $I=0$, the temperature stepwise up
to the value $k_BT=0.3 \Omega$ and solve at each step the non-linear problem
iteratively using a Newton-Raphson procedure. At this high temperature 
all quantum effects are smeared out, and we can replace the $B=0$ DOS
by the Gaussian Landau DOS corresponding to the required $\Omega$ value
without convergence problems.  Now we  raise
stepwise the current until the required value is reached. This calculation is
equivalent to the solution of the Drude problem discussed above. When
self-consistency is achieved, we lower the temperature stepwise until the  
required (low) value is reached. In each step we iterate until full
self-consistency is achieved, using the previous potential profile and the
conductivity tensor with the density profile of the previous step as starting
conditions. 

\begin{figure}[h]  \centering
\noindent  
\includegraphics[width=1.0\linewidth]{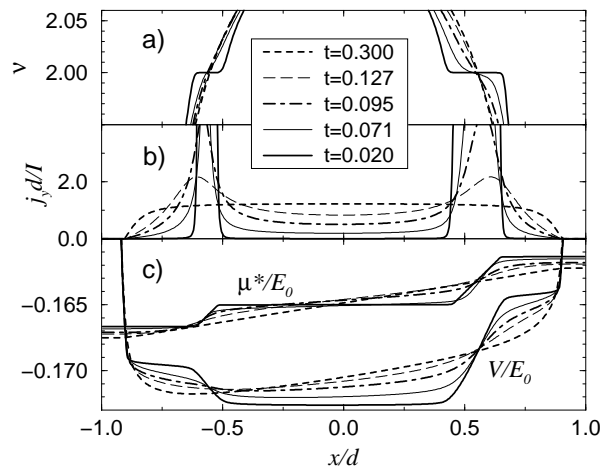}
\caption{\small Self-consistent results for a) filling factor $\nu(x)$, b)
  current density $j_y(x)$, and c) electrostatic and
  electrochemical potentials, $V(x)$ and $\mu^{\star}(x)$, at
  five temperatures $t=k_BT/ \hbar \omega_c$,  calculated from the
   Gaussian model.  ($\alpha=0.02$, $\hbar
  \omega_c/E_0=0.0053$, $\Gamma / \hbar \omega_c=0.03$, $U_H=\hbar \omega_c$). 
\label{fig:tempdevel}}
\end{figure}
Figure \ref{fig:tempdevel} shows the self-consistent results for several
intermediate temperatures. At the highest temperature (dashed lines) one
observes Drude-like 
behavior: the current density [Fig.~\ref{fig:tempdevel}(b)] is proportional to
the electron density [note that Fig.~\ref{fig:tempdevel}(a) shows the latter
only near local filling factor $\nu(x)=2$, while $\nu(0)=2.25$] and the
electrostatical and  
electrochemical potential increase nearly linearly across the 2DEG. With
decreasing 
temperature the 2DEG develops incompressible strips with low longitudinal
resistivity and the current density is increasingly confined to the
incompressible regions. Simultaneously the potentials develop a steplike
behavior with variation across the incompressible strips and plateaus in the
compressible regions.

\begin{figure}[h]  \centering
\noindent  
\includegraphics[width=1.05\linewidth]{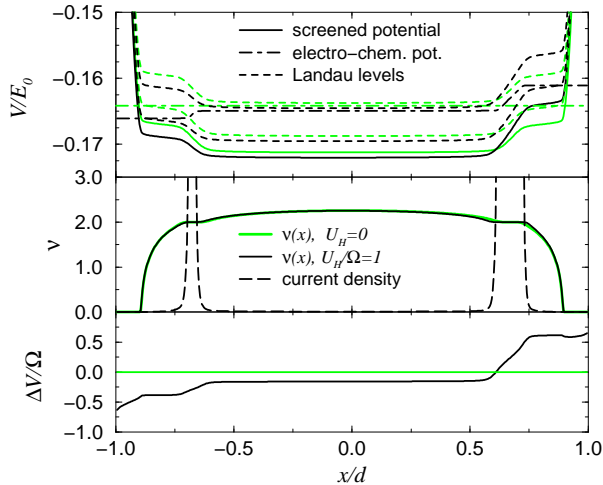}
\caption{\small As Fig.~\protect\ref{fig:flgsimpl}, but with $\sigma_l$
  calculated from the Gaussian model (\protect\ref{gauss-sigl}) instead of
  Eq.~(\protect\ref{gross_l}), for $U_H \equiv \Delta \mu^*= \Omega$. ($\Gamma
  / \hbar \omega_c=0.03$, all
  other parameters  as in  Fig.~\protect\ref{fig:flgsimpl}).
\label{fig:gauss005}}
\end{figure}

To evaluate the current-induced electrostatic potential $\Delta V$, we
perform the self-consistent calculation with and without applied current and
define  $\Delta V(x)=V(x;I,B,T)-V(x;0,B,T)$.  A
typical result is shown in Fig.~\ref{fig:gauss005}.  
The main difference between this result and Fig.~\ref{fig:flgsimpl} is that
now the current density is confined more strictly to a narrow region along the
incompressible strips (see dashed lines in the middle panels of the figures).
The more rapid decrease of the current density from the large values in the
incompressible strips to the small values in the compressible regions is
caused mainly by the much steeper increase of $\sigma_l(\nu)$ with increasing
$|\nu-2|$ [see Fig.~\ref{fig:slongvsnu}(b)]. The ratio between the values of
the current density in the compressible region and those in the incompressible
strips is also smaller, since we used near $\nu=2$ a smaller cutoff 
$\epsilon=\min[\sigma_l(\nu)/\sigma_H(\nu)]$  in the Gaussian model
($\epsilon= 10^{-4}$) than in  model (\ref{gross_l}) ($\epsilon =2\times
10^{-3}$). 
As a consequence, the variation of the electrochemical potential (dash-dotted
line of upper panel) and of the current-induced electrostatic potential (lower
panel) is practically confined to the region of the incompressible strips.

\begin{figure}[h]  \centering
\noindent  
\includegraphics[width=0.9\linewidth]{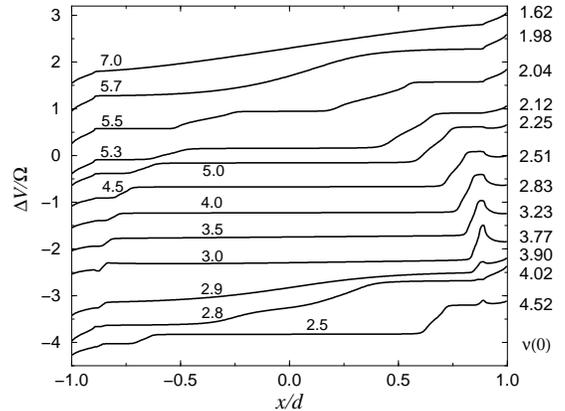}
\caption{\small Current-induced part $\Delta V(x)$ of the self-con\-sistent\-ly
  calculated   electrostatic potential in units of  the cyclotron energy
  $\Omega=\hbar\omega_c$, for several values of 
  $\Omega$. The numbers in the figure indicate $(\Omega/E_0)\times
  10^3$, those on the right hand side the corresponding values of
  $\nu(0)$. For clarity, the traces are shifted vertically by an arbitrary
  amount. 
\label{fig:ggdelvh_fuell}}
\end{figure}

The width and position of the incompressible strips and thus the locations of
strong variation of the current-induced potential change strongly with
varying magnetic field, i.e., with varying filling factors of the Landau
levels.
 In Fig.~\ref{fig:ggdelvh_fuell} we show results for selected values of
the magnetic field, leading to filling factors in the center of the
Hall bar that vary between $\nu(0)=1.62$ and $\nu(0)=4.52$. The temperature is
always chosen so low that the incompressible strips are well developed
($k_BT/\hbar \omega_c \lesssim 0.04$).

These results are easily understood. For $\hbar\omega_c\gtrsim 5.7\times
10^{-3}E_0$ no incompressible strips exist,  $\nu(x) <2$ for all $|x|<d$, and
the current density is largest near the center of the sample, where the
filling factor is 
largest and the longitudinal resistivity ($\rho_l\propto \sigma_l$) is
smallest. Therefore the gradient of $\Delta V(x)$ is largest in the center of
the Hall bar. If $\nu(0)<2$ is very close to 2,  $\rho_l(x=0)$ is very small,
the current density has a sharp maximum in the center, and the potential
profile has a strongly non-linear appearance (``type II'' behavior).
If $\nu(0)$ becomes considerably smaller than 2, $\rho_l(x)$ has
a broad maximum near $x=0$ and the current density profile follows essentially
the density profile, similar to the Drude case. This leads to an essentially
linear potential profile (``type I''), as is seen in the top curve of
Fig.~\ref{fig:ggdelvh_fuell} for $\hbar\omega_c=7.0\times 10^{-3}E_0$, with
$\nu(0)=1.62$. 
For slightly lower magnetic field, an incompressible strip with
$\nu(0)=2$ and (nearly) vanishing $\rho_l$ occurs in the center. Then the
variation of  $\Delta V(x)$ is 
confined essentially to this strip (type II). With still decreasing
$\hbar\omega_c$ this 
strip splits into two, which move with decreasing magnetic field towards to
edges of the Hall bar. The electrochemical potential and (apart from some
minor edge effects) the current-induced electrostatic potential then drop only
across these incompressible strips (``type III''), as is seen in
Fig.~\ref{fig:ggdelvh_fuell} for the curves with $2.04 \leq \nu(0) \leq 3.77$. 
As with further decreasing magnetic field the filling factor in the center
region comes close to 4 and thus $\rho_l(x=0)$ becomes small, a considerable
part of the current flows through this center 
region. Since at the same time the strips with local filling factor 2 become
very narrow, for $\nu(0) \lesssim 4$ a considerable part of the induced
potential drops in a broad center region (type II). For $\nu(0) >4$ the center
region becomes again compressible, with constant $\Delta V(x)$, and the
incompressible strips with $\nu(x)=4$, across which now most of the Hall
voltage drops,  move away from the center (type III). The lowest trace in
Fig.~\ref{fig:ggdelvh_fuell} shows such a situation with tiny structures at
the edges of the electron density profile which are due to the outer
incompressible strips with local filling factor 2.

\section{Conclusions} \label{conclusions}
Our results for the current-induced Hall-potential profile
(Fig.~\ref{fig:ggdelvh_fuell}) reproduce characteristic features of the
experiment of Ahlswede {\em et al.} \cite{Ahlswede01:562}. If the filling
factors $\nu(0)$ in the center of the sample are close to, but below 
integer values, the potential drops in a non-linear fashion in a broad center
region. For $\nu(0)$ values slightly larger than the integer values, the
potential is constant in the center region and drops exclusively across the
incompressible strips. Of course, our results show this characteristic
behavior only near even integer values of $\nu(0)$, since we have neglected
spin-splitting, whereas in the experiment spin-splitting is resolved and this
behavior occurs also near small odd-integer values of $\nu(0)$.

This characteristic dependence of the Hall-potential profile on the magnetic
field cannot be explained by the previous calculations assuming
dissipationless Hall currents \cite{Pfannkuche92:7032,Oh97:13519,Oh97:108}
and emphasizes the importance of dissipation. From the nice qualitative
agreement of our results with the experimental data we conclude that our local
equilibrium approach, which combines dissipative transport with screening
effects and
allows to calculate electron and current density as well as electrostatic and
electrochemical potential self-consistently, contains most of the relevant
physics. There is, however, room and need for improvements. 

One desirable improvement concerns the effectiveness of narrow incompressible
strips. To avoid numerical divergencies, we used a cut-off
$\epsilon\cdot\sigma_H$ 
for $\sigma_l$ at even integer values of $\nu$, with $\epsilon=10^{-4}$. If we
take the limit $\epsilon \rightarrow 0$ and sufficiently low temperatures,
$\rho_l(x)$ becomes exponentially small in the incompressible strips. Then,
according to Eq.~(\ref{normalize}), the electric field along the Hall bar and,
therefore, the longitudinal resistance become exponentially small whenever an
incompressible strip exists, and not only for a limited interval of magnetic
field values in a plateau region of the QHE. 
To eliminate this unreasonable
behavior, we should include a mechanism (other than the simple cutoff) that
limits the current density in, and thereby the voltage drop across, narrow
incompressible strips. Such a mechanism could make the incompressible strips
with local filling factor $\nu(x)=2$ ineffective for magnetic fields with
$\nu(0)\gtrsim 3$. This would turn the ``type III'' curves with $\nu(0)=3.23$
and 3.77 in Fig.~\ref{fig:ggdelvh_fuell} into quasi-linear ``type I'' curves
and would eliminate the tiny edge-near structures in the three lowest curves.
Both changes would improve the agreement with the experiment.

Several physical effects may lead to such a mechanism. One is the non-local
relation between the current density and the driving electric field, which we
have approximated by a strictly local one. Another one is Joule heating, which
is most effective where the current density is high and may destroy narrow
incompressible strips, i.e., lead to a local breakdown of the QHE.
A systematic treatment of heating effects will require the consideration of
energy balance and heat conduction, as has recently been pointed out by Akera
\cite{Akera01:1468}
in his hydrodynamic approach to  quantum Hall systems in the breakdown regime.
A consideration of the  heating processes relevant under QHE conditions
 (e.g.\ ``quasi-elastic inter-Landau-level scattering'')
 \cite{Eaves86:346,Akera00:3174} may even demand a treatment beyond
 a local hydrodynamic approximation and  require a more microscopic non-local
 description of  stationary current-carrying 
non-equilibrium states intermediate between the zero-resistance quantum Hall
state and the finite-resistance breakdown state.\cite{Guven02:155316} 
Such a microscopic approach to heating and resistive processes may also open
the possibility of a unified description of dissipative currents, which we
have considered phenomenologically in the present paper, and non-dissipative
equilibrium currents, which we have mentioned in the introduction but
completely neglected in the calculations.

Finally we want to mention that in our model calculations
the imposed current leads to a broadening of
the incompressible strips on one side of the sample and to a narrowing of the
corresponding strips on the opposite side (of course the strips exchange their
role if we invert the direction of the current). This asymmetry is clearly
seen in Fig.~\ref{fig:ggdelvh_fuell}, and can also be observed in Fig.~1 of
Ref.~\onlinecite{Ahlswede01:562}. A systematic investigation of this effect
may be of interest.

\acknowledgments
We gratefully acknowledge helpful discussions with E. Ahlswede, G. Nachtwei,
and J. Weis, and financial support by the Deutsche Forschungsgemeinschaft, SP
``Quanten-Hall-Systeme'' GE306/4-1.


\begin{thebibliography}{10}

\bibitem{Komiyama96:2067}
S. Komiyama and H. Hirai, Phys. Rev. B {\bf 54},  2067  (1996).

\bibitem{Wexler94:4815}
C. Wexler and D.~J. Thouless, Phys. Rev. B {\bf 49},  4815  (1994).

\bibitem{Geller94:11714}
M.~R. Geller and G. Vignale, Phys. Rev. B {\bf 50},  11714  (1994).

\bibitem{Geller95:14137}
M.~R. Geller and G. Vignale, Phys. Rev. B {\bf 52},  14137  (1995).

\bibitem{Balaban95:R5503}
N.~Q. Balaban, U. Meirav, and H. Shtrikman, Phys. Rev. B {\bf 52},  R5503
  (1995).

\bibitem{Yahel97:537}
E. Yahel, A. Palevski, and H. Shtrikman, Superlattices and Microstructures {\bf
  22},  537  (1997).

\bibitem{Nachtwei97:6731}
G. Nachtwei, G. L{\"u}tjering, D. Weiss, Z. Liu, K. von Klitzing, and C. Foxon,
  Phys. Rev. B {\bf 55},  6731  (1997).

\bibitem{Nachtwei98:9937}
G. Nachtwei, Z.~H. Liu, G. L{\"u}tjering, R.~R. Gerhardts, D. Weiss, K. von
  Klitzing, and K. Eberl, Phys. Rev. B {\bf 57},  9937  (1998).

\bibitem{Weitz00:247}
P. Weitz, E. Ahlswede, J. Weis, K. v.~Klitzing, and K. Eberl, Physica E {\bf
  6},  247  (2000).

\bibitem{Ahlswede01:562}
E. Ahlswede, P. Weitz, J. Weis, K. v.~Klitzing, and K. Eberl, Physica B {\bf
  298},  562  (2001).

\bibitem{Ahlswede02:165}
E. Ahlswede, J. Weis, K. v.~Klitzing, and K. Eberl, Physica E {\bf 12},  165
  (2002).

\bibitem{Weitz00:349}
P. Weitz, E. Ahlswede, J. Weis, K. v.~Klitzing, and K. Eberl, Appl. Surf. Sci.
  {\bf 157},  349  (2000).

\bibitem{Chklovskii92:4026}
D.~B. Chklovskii, B.~I. Shklovskii, and L.~I. Glazman, Phys. Rev. B {\bf 46},
  4026  (1992).

\bibitem{Chklovskii93:12605}
D.~B. Chklovskii, K.~A. Matveev, and B.~I. Shklovskii, Phys. Rev. B {\bf 47},
  12605  (1993).

\bibitem{Lier94:7757}
K. Lier and R.~R. Gerhardts, Phys. Rev. B {\bf 50},  7757  (1994).

\bibitem{Chang90:871}
A.~M. Chang, Solid State Commun. {\bf 74},  871  (1990).

\bibitem{Pfannkuche92:7032}
D. Pfannkuche and J. Hajdu, Phys. Rev. B {\bf 46},  7032  (1992).

\bibitem{Oh97:13519}
J.~H. Oh and R.~R. Gerhardts, Phys. Rev. B {\bf 56},  13519  (1997).

\bibitem{Woltjer86:149}
R. Woltjer, R. Eppenga, J. Mooren, C.~E. Timmering, and J.~P. Andr{\'e},
  Europhys. Lett. {\bf 2},  149  (1986).

\bibitem{Woltjer87:104}
R. Woltjer, R. Eppenga, and M.~F.~H. Schuurmans,  in {\em High Magnetic Fields
  in Semiconductor Physics}, Vol.~71 of {\em Springer Series in Solid-State
  Sciences}, edited by G. Landwehr (Springer-Verlag, Berlin, 1987), p.\ 104.

\bibitem{Gerhardts99:2561}
R.~R. Gerhardts and J. Gro{\ss }, Phys. Rev. B {\bf 60},  2561  (1999).

\bibitem{Guven02:Oxford}
K. G{\"u}ven and R.~R. Gerhardts,  in {\em Proc. 15th Intern. Conf. on High
  Magnetic Fields in Semicond. Phys.} (Oxford, UK, 2002), unpublished.

\bibitem{Glazman91:32}
L.~I. Glazman and I.~A. Larkin, Semicond. Sci. Technol. {\bf 6},  32  (1991).

\bibitem{Gossmann98:1680}
U.~J. Gossmann, A. Manolescu, and R.~R. Gerhardts, Phys. Rev. B {\bf 57},  1680
   (1998).

\bibitem{Oh97:108}
J.~H. Oh and R.~R. Gerhardts, Physica E {\bf 1},  108  (1997).

\bibitem{Gross98:60}
J. Gro{\ss } and R.~R. Gerhardts, Physica B {\bf 256-258},  60  (1998).

\bibitem{Gerhardts75:285}
R.~R. Gerhardts, Z. Physik B {\bf 21},  285  (1975).

\bibitem{Ando82:437}
T. Ando, A.~B. Fowler, and F. Stern, Rev. Mod. Phys. {\bf 54},  437  (1982).

\bibitem{Akera01:1468}
H. Akera, J. Phys. Soc. Jpn. {\bf 70},  1468  (2001).

\bibitem{Eaves86:346}
L. Eaves and F.~W. Sheard, Semicond. Sci. Technol. {\bf 1},  346  (1986).

\bibitem{Akera00:3174}
H. Akera, J. Phys. Soc. Jpn. {\bf 69},  3174  (2000).

\bibitem{Guven02:155316}
K. G{\"u}ven, R.~R. Gerhardts, I.~I. Kaya, B.~E. Sagol, and G. Nachtwei, Phys.
  Rev. B {\bf 65},  155316  (2002).

\end{thebibliography}


\end{document}